# Coulomb-Modified Fano Resonance in a One-Lead Quantum Dot

A. C. Johnson, C. M. Marcus
*Department of Physics, Harvard University, Cambridge, Massachusetts 02138*
M. P. Hanson, A. C. Gossard
*Department of Materials, University of California, Santa Barbara, California 93106*

We investigate a tunable Fano interferometer consisting of a quantum dot coupled via tunneling to a one-dimensional channel. In addition to Fano resonance, the channel shows strong Coulomb response to the dot, with a single electron modulating channel conductance by factors of up to 100. Where these effects coexist, lineshapes with up to four extrema are found. A model of Coulomb-modified Fano resonance is developed and gives excellent agreement with experiment.

The interplay between interference and interaction, in its many forms, is the central problem in mesoscopic physics. In bulk systems, screening reduces the strong repulsion between electrons to a weak interaction between quasiparticles, but in confined geometries electron-electron interaction can dominate transport. The Fano effect—an interference between resonant and non-resonant processes—was first proposed in the context of atomic physics [1]. More recently, Fano resonances have been investigated in condensed matter systems, including surface impurities [2], quantum dots [3, 4], and carbon nanotubes [5], and have generated interest as probes of phase coherence [6, 7] and as possible spin filters [8]. These studies have treated Fano resonance as purely an interference effect. When the resonant channel is a tunneling quantum dot, however, Fano resonance coexists with Coulomb interaction that appears in single-dot transmission as the Coulomb blockade. In the Fano regime, the coexistence of Coulomb and interference effects leads to new transport regimes that to date have not been investigated theoretically or experimentally.

In this Letter, we present measurements of a Fano interferometer consisting of a quantum dot coupled to a one-dimensional channel. Independent control of all couplings controlling both resonant and non-resonant processes allows several regimes to be investigated: For instance, when the channel is partially transmitting and there is *no* tunneling into the dot, a charge-sensing regime is observed in which the channel conductance responds to the number of charges on the dot [9]. At the other extreme, when the dot-channel barrier is lowered and the channel pushed toward the dot, such that *all* transmitted electrons must pass through the dot, standard Coulomb blockade resonances are observed. Between these limiting cases lies the Coulomb-modified Fano regime, in which resonant tunneling via the dot and direct channel transmission coexist. Fano resonances appear throughout this regime, generally in conjunction with charge sensing. A model that combines these two effects is developed below, allowing resonance parameters to be extracted and interaction effects to be evaluated.

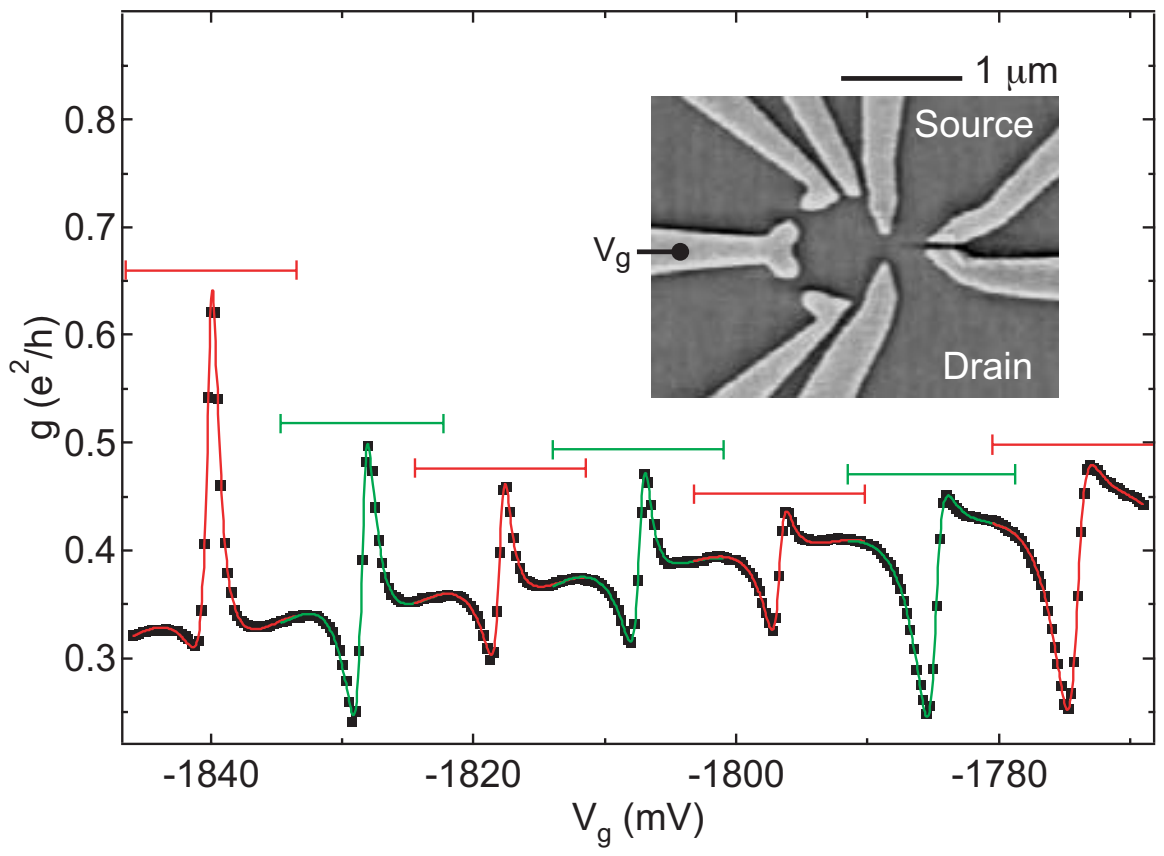

FIG. 1 Channel conductance data (squares) and fits (curves) vs. gate voltage $V_g$ in the Fano regime. Bars show fitting ranges. Inset: SEM image of a similar sample, a quantum dot coupled by one lead to a conducting channel.

The device (inset to Fig. 1) we consider consists of a 0.5 $\mu m^2$ quantum dot with several adjacent gates forming a channel, defined by Cr-Au depletion gates on a $GaAs/Al_{0.3}Ga_{0.7}As$ heterostructure grown by MBE. The two-dimensional electron gas lies 100 nm below the surface, with density $2 \times 10^{11}$ $cm^{-2}$ and mobility $2 \times 10^5$ $cm^2/Vs$. The dot contains ~ 600 electrons and has level spacing $\Delta \sim 20$ μeV. The experiment was performed in a dilution refrigerator with an electron temperature of 50 mK, in a magnetic field of 0 to 200 mT perpendicular to the device plane. Conductance was measured using a lock-in amplifier with 10 μV excitation at 15.7 Hz.

Figure 1 illustrates a typical trace of conductance as a function of gate voltage $V_g$ , over a range containing seven resonances, in the Coulomb-modified Fano regime. A progression of lineshapes is seen, each comprising a dip and a peak similar to a simple Fano form. However, the non-interacting Fano lineshape, given by Eq. (1) below, is insufficient to explain these data. However, a model incorporating Coulomb interaction (described below) provides improved fits, as seen in Fig. 1.

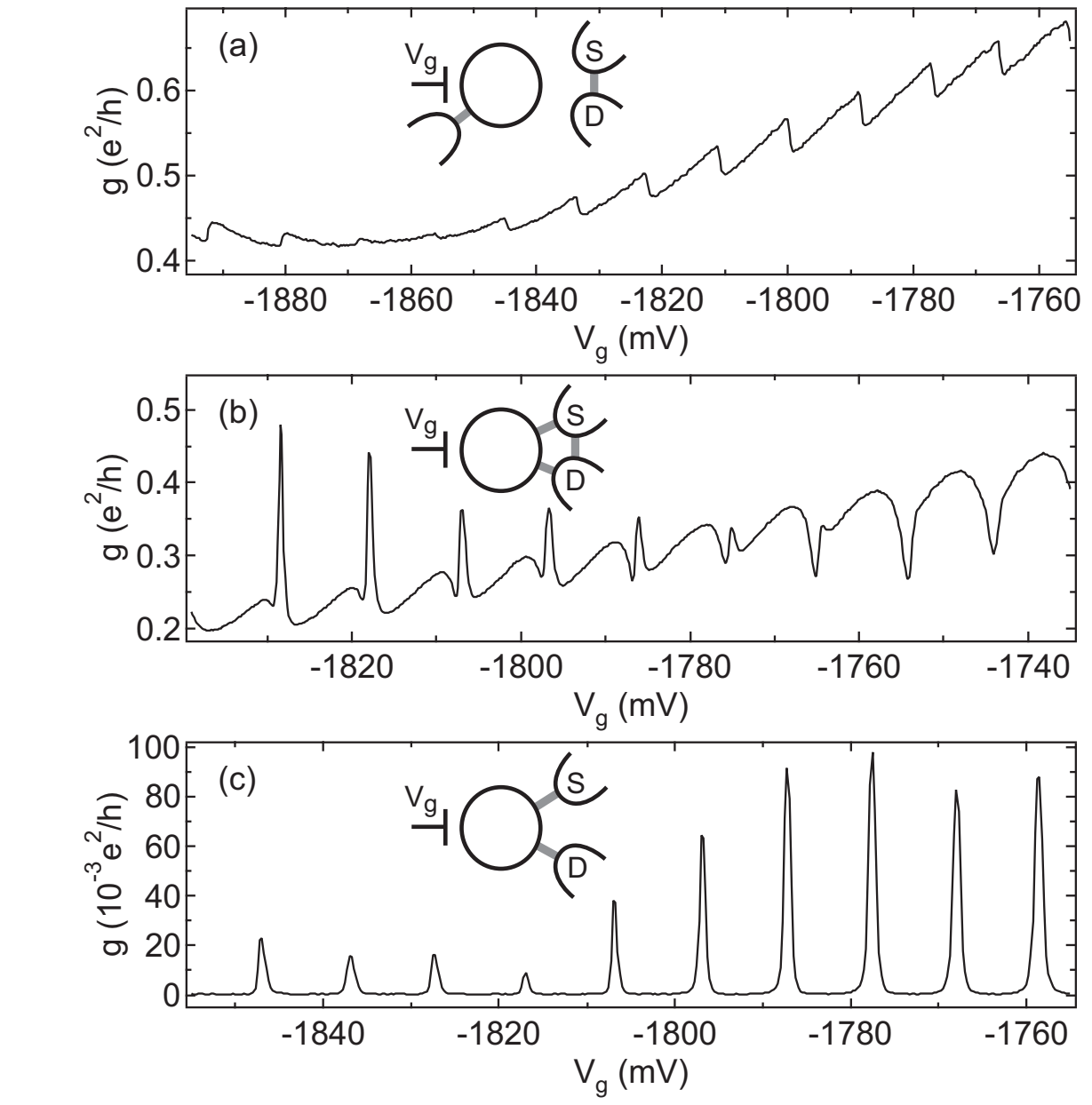

FIG. 2 Three configurations with a tunnel coupled dot. Drawings indicate tunneling paths. (a) Pure charge sensing: the dot couples capacitively to the channel and tunnels weakly to a third reservoir. (b) Fano resonance with charge sensing: tunneling between the channel and the dot interferes with direct transmission. (c) Breit-Wigner resonances: the only conducting path is through the dot.

We first discuss qualitatively the limiting cases and intermediate Coulomb-modified Fano regime mentioned above. When the channel is tuned to partially transmit one mode, but there is no conductance between channel and dot, sawtooth patterns as in Fig. 2(a) appear. This can be understood in terms of a single *effective* gate, combining the influence of the metal gate and the dot charge, that smoothly modulates the channel conductance, $g$. A single electron on the dot contributes to the effective gating of the channel an amount $V_s$, denoted the *sensing voltage*. Each time a charge is added to the dot (as $V_g$ is made more positive), the effective gate voltage jumps by $V_s$, causing $g$ to jump to the value it had at a gate voltage more negative by $V_s$. For this device, $V_s$ is typically ~80% of the spacing between jumps, indicating that the channel is more sensitive to excess dot charge than to the gate directly.

The Coulomb-modified Fano regime emerges as tunneling is introduced between the dot and the channel. When the charge sensing effect is relatively weak, such as in Fig. 1, features clearly resembling non-interacting Fano resonance are observed. In other cases, for example Fig. 2(b), the charge-sensing jump is comparable to or larger than the peak of the Fano resonance. In this regime, combining the charge-sensing jump with the dip-peak pair of Fano resonance, one resonance can have up to four extrema.

When direct conductance is made much smaller than conductance through the dot, the direct path (through the channel) no longer interferes significantly with the resonant path (though the dot). In this limit, the Fano regime crosses over to the familiar Coulomb blockade regime, yielding simple single resonances as seen in Fig. 2(c).

Before analyzing the Fano regime in detail, we turn our attention to features that can appear as Fano resonance but are actually a result of charge sensing. Figure 3 shows the effect of quantized charge in the dot on a resonance in the channel, in the absence of any tunneling between the two. Channel conductance was measured while tuning the coupling between the dot and a third reservoir. When the tunnel rate to the third reservoir is low (Fig. 3(a)), one finds smooth segments punctuated by jumps. The dotted curves are identical forms offset by $V_s = 4.0$ mV, indicating how channel resonances would appear for an isolated dot occupied by $n = N$ or $n = N+1$ electrons. If this family of curves is extended to all $n$, it overlays every segment of the data, though the jumps from one curve to the next are irregularly spaced and change position if the sweep is repeated. We identify these jumps as single tunneling events, and estimate a tunneling time $\hbar/\Gamma$ here of order seconds, based on motion of the jump point upon repeating sweeps.

As tunneling to the third reservoir is increased (Fig. 3(b)), jumps become periodic and repeatable, but there remains a family of evenly spaced curves onto which all of the data falls. In this case the channel resonance is much narrower, and adding a single charge to the dot shifts the channel from directly on to far off resonance. This single-electron switch has an on/off conductance ratio of 20. Traces with still narrower resonances show on/off ratios >100, as in other recent reports [10]. Notice, however, the similarity between these asymmetric line shapes and Fano resonances, even though this is pure charge sensing.

Still greater coupling of the dot to the third reservoir yields lifetime-broadened transitions. This is the case in Fig. 3(c), where the dominant feature is a single broad resonance, modulated by weak charge quantization in the dot. This motivates an important feature of the model, that the Fano resonance and the charge-sensing jump, since both result from the same process, have a single width parameter $\Gamma$.

In the single-level transport regime, $kT < \Gamma < \Delta$, transmission through one discrete level produces a Breit-Wigner resonance, represented by a complex transmission amplitude $t = t_0/(\varepsilon + i)$ with dimensionless detuning $\varepsilon = (E - E_0)/(\Gamma/2)$. Here an electron of energy $E$ encounters a resonance at $E_0$ with width $\Gamma$, and peak transmission $t_0$, accounting for lead asymmetry. Conductance is proportional to the transmission probability $|t|^2$, giving a Lorentzian lineshape [11]. A continuum channel can be added to the amplitude (coherently) or probability (incoherently), giving the non-interacting Fano lineshape,

$$g(E) = g_{inc} + g_{coh}\frac{|\varepsilon + q|^2}{\varepsilon^2 + 1}, \qquad (1)$$

where $g_{coh}$ ($g_{inc}$) is the coherent (incoherent) contribution to the continuum conductance. The Fano parameter $q$ selects from a symmetric peak ($q = \infty$), symmetric dip ($q = 0$), or a dip to the left ($q > 0$) or right ($q < 0$) of a peak. The resonances in Fig. 1 evolve from $q = 2.5$ on the left to $q = 0.6$ on the right. In cases allowing arbitrary phase between the resonant and non-resonant paths, such as the Aharonov-Bohm interferometer, the $q$ becomes complex. This is equivalent to increasing $g_{inc}$. In the present context, the interfering paths enclose no area, forcing a real $q$. This eliminates the ambiguity between $g_{inc}$ and Im$q$, and constrains the Fano lineshape to maximum visibility [6].

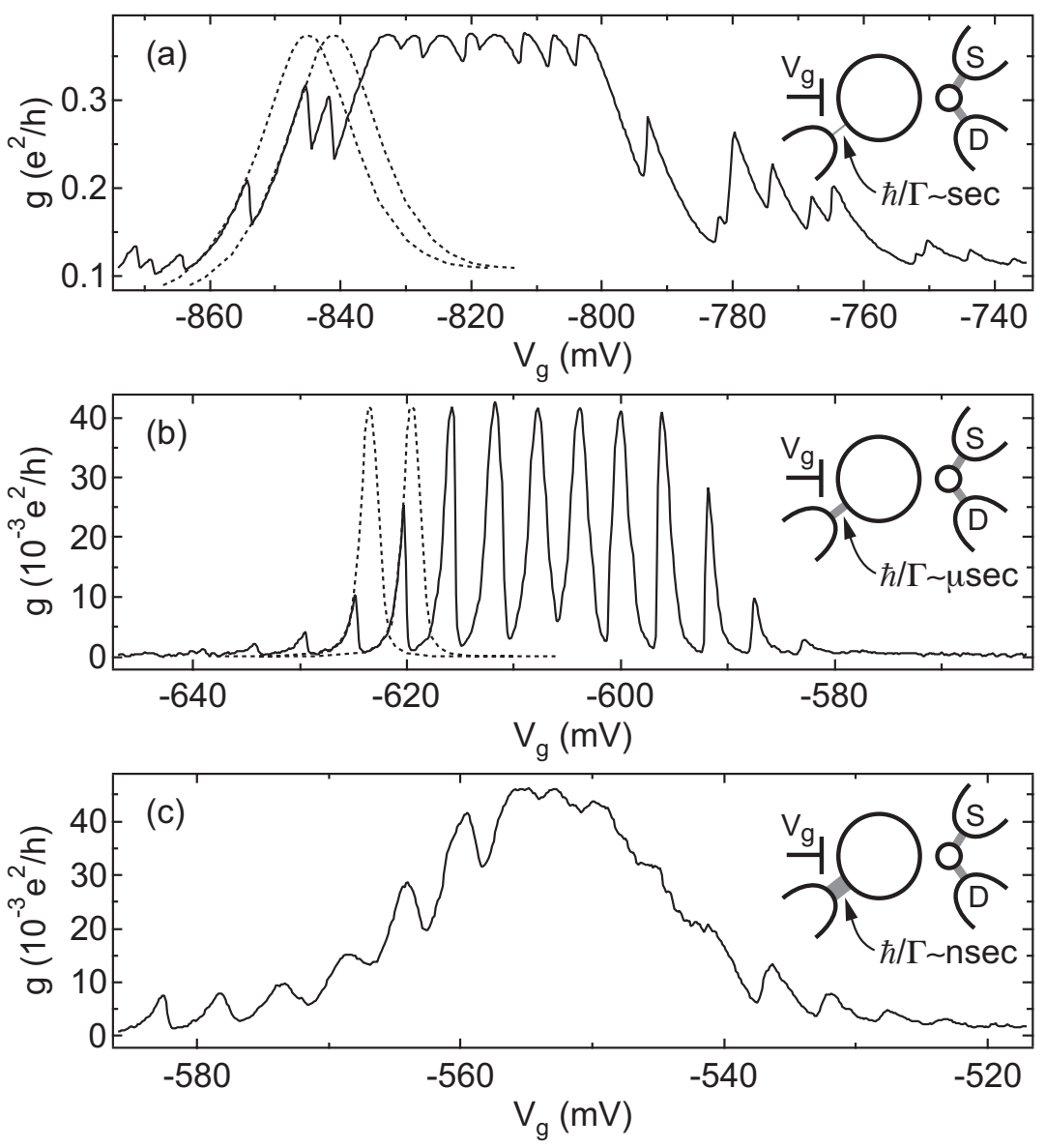

FIG. 3 Charge sensing by a channel resonance. Drawings indicate the tunneling rate to the extra lead. Dotted curves show how the full resonance would look with dot charge fixed at two consecutive values. (a) With the dot nearly isolated, jumps between these curves are sharp, erratic and unrepeatable. (b) Increase tunneling and the jumps become periodic and repeatable. (c) With the dot nearly open, the jumps broaden and resemble an oscillation superimposed on a single, broad resonance.

The Coulomb-modified Fano effect can be modeled by extending Eq. (1) to include nearby resonances and charge sensing effects. We first write $g$ as a sum over initial occupation numbers of the dot,

$$g(V_g) = \sum_n g_n(V_g) p_n(V_g). \qquad (2)$$

If the dot contains $n$ electrons, the only resonant processes allowed are the $n \rightarrow n+1$ and $n-1 \rightarrow n$ transitions, corresponding to the addition or removal of one electron. The contribution of the $n$-electron dot is thus

$$g_n(V_g) = g_{inc}(\tilde{V}_n) + g_{coh}(\tilde{V}_n)\left|1 + \frac{q(\tilde{V}_n) - i}{\varepsilon_{n-} + i} + \frac{q(\tilde{V}_n) - i}{\varepsilon_{n+} + i}\right|^2. \qquad (3)$$

The allowed resonances have detunings $\varepsilon_{n\pm} = (eV_gC_g/C_{tot} + E_{n\rightarrow n\pm1})/(\Gamma/2)$, including a contribution from $V_g$ (in energy units), with a lever arm given by the ratio of gate capacitance to total dot capacitance. The allowed resonances add coherently to a direct conductance $g_{coh}$ [12]; $g_{inc}$ is then added to account for multiple modes in the channel and explicit decoherence processes. Finally, each term is weighted by $p_n(V_g) = [\tan^{-1}(\varepsilon_{n-}) - \tan^{-1}(\varepsilon_{n+})]/\pi$, the zero-temperature probability of occupation $n$. This can be derived from the Friedel sum rule, which relates changes $\delta(\ldots)$ in transmission phase to fractional changes in dot occupancy, $\delta(\arg(t)) = \pi\,\delta(\langle N\rangle) \bmod \pi$, where $t$ is the Breit-Wigner transmission or reflection amplitude [13, 14].

Charge sensing enters Eq. (3) via the effective gate voltage $\tilde{V}_n = V_g - nV_s$. We expand dependences on $\tilde{V}_n$ to first order, $q(\tilde{V}_n) = q_0 + \tilde{V}_n dq/dV_g$, and similarly for $g_{coh}$ and $g_{tot} = g_{inc} + g_{coh}$. The slope $dg_{tot}/dV_g$ gives the charge-sensing sawtooth pattern, where charges in the dot affect the potential seen by charges in the channel. Nonzero $dq/dV_g$ and $dg_{coh}/dV_g$ produce subtle changes to the lineshape near resonance, and reveal how the potential seen by a charge in resonance depends self-consistently on its probability of being in the dot. As nearby resonances are generally similar in lineshape, the model assumes they obey the same linear expressions when calculating their influence on the tails of the resonance being fitted. This permits fitting with overlapping ranges as shown in Fig. 1, for more accurate determination of the charge-sensing parameters.

One limitation of this model deserves particular attention. A resonance in the channel makes conductance away from a dot resonance nonlinear on the scale of $V_s$, while the model assumes linearity. Thus, while the model trivially fits Fig. 2(a) by setting $g_{coh} = 0$, it cannot account for Fig. 3 in this manner. However, the asymmetric lineshapes in Fig. 3 give a somewhat plausible fit to the model. In order to unambiguously identify a Fano resonance, it is necessary that the off-resonant conductance varies linearly, which in turn requires that resonances be well separated so that background behavior can be isolated.

Figure 4 shows fits to the model and the information this yields. Every resonance in Fig. 4(a) was fit four times, using all permutations of varying or fixing at zero the parameters $dq/dV_g$ and $d\,g_{coh}/dV_g$, to investigate whether this unusual sensing is necessary to explain the data. The results for one resonance are shown in Fig. 4(b). The first fit holds both parameters at zero while $dg_{tot}/dV_g$, $V_S$ and the five parameters of a basic Fano resonance are varied. This reproduces all features qualitatively, but quantitative agreement is much poorer than with the latter three fits, which are nearly identical and agree with experiment to almost within the noise [15]. We therefore conclude that at least one of $q$ and $g_{coh}$ is subject to charge sensing, implying that the potential felt by the charge in resonance is modified by its own probability to be in the dot.

Finally, we consider parameter correlations among subsequent resonances. Figures 4(c)-(e) show $g_{tot}$ and $g_{coh}$, $\Gamma$, and $q$ extracted from each fit to each resonance in Fig. 4(a). Most trends are consistent with general arguments about tunneling wave functions: as $g_{tot}$ increases, indicating a lower channel potential, $\Gamma$ increases due to a lower tunnel barrier, and $q$ decreases to keep peak conductance, given by $g_{inc} + g_{coh}(q^2+1)$, relatively constant. On top of this there appear to be fluctuations in $g_{coh}$, $\Gamma$, and $q$ which are expected due to subsequent dot wave functions having different amplitudes near the barrier.

Two additional features stand out in the data. First, in many instances the fractional coherence $g_{coh}/g_{tot}$ is roughly

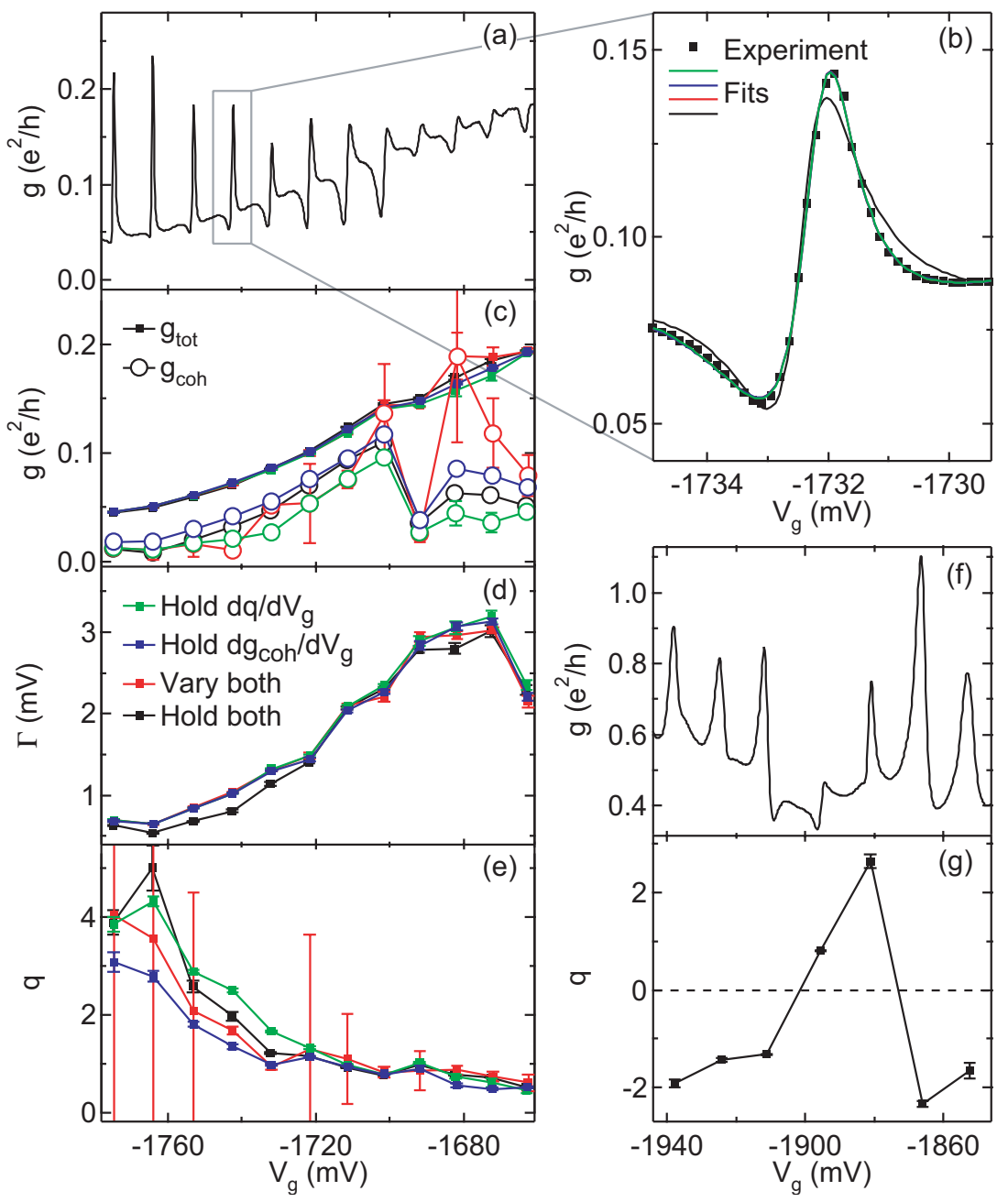

FIG. 4 (a) Experimental data with twelve Fano resonances. In (b) we show one resonance and its four fits. Of the parameters dq/dVg and dgcoh/dVg, the fits fix at zero both (black), one (green and blue), or neither (red). All four fits are shown, but the latter three are indistinguishable. In (c)-(e) we plot, using the same colors to denote fitting method, the parameters $g_{tot}$ and $g_{coh}$, G, and q, from the fits to (a). Panel (f) shows data exhibiting reversals of the sign of q, with extracted q values shown in (g).

constant for several peaks then jumps abruptly to a different level for subsequent peaks, as occurs in Fig. 4 at $V_g = -1700$ mV, while $g_{tot}$ and other parameters evolve smoothly. Observed fractional coherence spans a range from <10% to >50% [16], likely due to multiple weakly transmitting modes in the channel which couple differently to the dot. A jump in $g_{coh}/g_{tot}$ may reflect abrupt rearrangement of dot wave functions, changing its coupling to each channel mode while total coupling, measured by Γ, is nearly unchanged.

Second, changes to the sign of $q$ are present but infrequent. An example is the data in Fig. 4(f) and (g), where $q$ flips twice in the observed range. Previous observations of mesoscopic Fano resonance in transmission [3], including measurements of a dot in an Aharonov-Bohm ring [17] showed a constant sign of $q$, sparking intense debate on why no reversals were seen when simple theory predicts that consecutive peaks always change sign [14, 18]. In the present geometry, however, as with recent work on Fano resonance in reflection [4], a constant sign is expected because the dot meets the channel in only one lead, so there is no freedom of relative sign between two matrix elements to reverse the phase. Why then are different signs of $q$ observed here at all? One possibility is that the scattering phase in the channel changes by $\pi$, as if it too passed through a resonance [19]. This requires $q$ to pass smoothly through zero at a maximum of $g_{tot}$, which is consistent with the observation that $q$ is mostly positive in some regions and negative in others, but cannot explain the flips in Fig. 4(f). A more likely explanation for this data is that the source and drain leads couple to different areas in the dot, due to the spatial extent of the tunnel barrier. With appropriate nodes in several wave functions, the source and drain couplings have opposite signs and $q$ reverses sign for several resonances, exactly as observed in Fig. 4(f). In short, both coherence jumps and Fano parameter flips can be explained by imperfect one-dimensionality of the channel and the tunnel barrier. In principle, these effects could be used to study wave function properties not accessible via transmission.

We thank A. A. Clerk, W. Hofstetter and B. I. Halperin for useful discussion and N. J. Craig for experimental contributions. This work was supported by the ARO under DAAD19-99-1-0215 and DAAD55-98-1-0270, and NSF-NSEC (PHY-0117795). ACJ acknowledges support from the NSF Graduate Research Fellowship Program.

[1] U. Fano, Phys. Rev. **124**, 1866 (1961).

[2] V. Madhavan, *et al.*, Science **280**, 567 (1998).

[3] J. Gores, *et al.*, Phys. Rev. B **62**, 2188 (2000); I. G. Zacharia, *et al.*, Phys. Rev. B **64**, 155311 (2001); C. Fuhner, *et al.*, cond-mat/0307590 (2003); K. Kobayashi, *et al.*, Phys. Rev. Lett. **88**, 256806 (2002).

[4] K. Kobayashi, *et al.*, cond-mat/0311497 (2003).

[5] J. Kim, *et al.*, Phys. Rev. Lett. **90**, 166403 (2003).

[6] A. A. Clerk, X. Waintal, and P. W. Brouwer, Phys. Rev. Lett. **86**, 4636 (2001).

[7] Y.-J. Xiong and S.-J. Xiong, Int. J. Mod. Phys. B **16**, 1479 (2002).

[8] J. F. Song, Y. Ochiai, and J. P. Bird, Appl. Phys. Lett. **82**, 4561 (2003).

[9] M. Field, *et al.*, Phys. Rev. Lett. **70**, 1311 (1993); I. Amlani, *et al.*, Appl. Phys. Lett. **71**, 1730 (1997); D. S. Duncan, *et al.*, Appl. Phys. Lett. **74**, 1045 (1999).

[10] I. H. Chan, *et al.*, Appl. Phys. Lett. **80**, 1818 (2002); I. H. Chan, *et al.*, Physica E **17**, 584 (2003).

[11] G. Breit and E. Wigner, Phys. Rev. **49**, 519 (1936).

[12] Adding amplitudes is an approximation which inflates the extracted coherence for poorly-separated resonances. Going beyond this requires choosing a microscopic model. See A. Aharony, *et al.*, Phys. Rev. Lett. **66**, 115311 (2002).

[13] T. K. Ng and P. A. Lee, Phys. Rev. Lett. **61**, 1768 (1988).

[14] T. Taniguchi and M. Buttiker, Phys. Rev. B **60** (1999).

[15] In some cases all four fits are noticeably different, but a near-degeneracy is typical. Therefore, the fit is unreliable when both parameters are varied, hence the erratic behavior and large error bars on the parameters in red.

[16] This is a conservative bound from the ratio of minimum conductance to $g_{tot}$. Extracted $g_{coh}/g_{tot}$ reaches 90%, because charge sensing raises the minimum even if $g_{inc} = 0$.

[17] A. Yacoby, *et al.*, Phys. Rev. Lett. **74**, 4047 (1995); R. Schuster, *et al.*, Nature **385**, 417 (1997).

[18] R. Baltin and Y. Gefen, Phys. Rev. Lett. **83**, 5094 (1999); A. Levy Yeyati and M. Buttiker, Phys. Rev. B **62** (2000); H. Q. Xu and B.-Y. Gu, J. Phys. Cond. Mat. **13**, 3599 (2001); T.-S. Kim and S. Hershfield, Phys. Rev. B **67**, 235330 (2003).

[19] A. A. Clerk (private communication).